\documentclass[preprint,superscriptaddress,showpacs,preprintnumbers,amsmath,amssymb]{revtex4}

\usepackage{graphicx}
\usepackage{dcolumn}
\usepackage{bm}

\def\lesssim{\ \raise.3ex\hbox{$<$}\kern-0.8em\lower.7ex\hbox{$\sim$}\ }
\def\gesim{\ \raise.3ex\hbox{$>$}\kern-0.8em\lower.7ex\hbox{$\sim$}\ }

\def\rnum#1{\expandafter{\romannumeral #1}} 
\def\Rnum#1{\uppercase\expandafter{\romannumeral #1}} 

\begin{document}

\title{Pseudogap phenomenon in an ultracold Fermi gas with a p-wave pairing interaction}

\author{Daisuke Inotani$^1$, Ryota Watanabe$^1$, Manfred Sigrist$^2$, and Yoji Ohashi$^{1,3}$}
 \affiliation{$^1$Department of Physics, Keio University, 3-14-1 Hiyoshi, Kohoku-ku, Yokohama 223-8522, Japan \\
 $^2$Institut f{\"u}r Theoretische Physik, ETH Z{\"u}rich,CH-8093 Z{\"u}rich, Switzerland\\ 
$^3$CREST(JST), 4-1-8 Honcho, Saitama 332-0012, Japan}
\date{\today}

\begin{abstract}
We investigate single-particle properties of a one-component Fermi gas with a tunable $p$-wave interaction. Including pairing fluctuations associated with this anisotropic interaction within a $T$-matrix theory, we calculate the single-particle density of states, as well as the spectral weight, above the superfluid transition temperature $T_{\rm c}$. Starting from the weak-coupling regime, we show that the so-called pseudogap first develops in these quantities with increasing the interaction strength. However, when the interaction becomes strong to some extent, the pseudogap becomes obscure to eventually disappear in the strong-coupling regime. This non-monotonic interaction dependence is quite different from the case of an $s$-wave interaction, where the pseudogap simply develops with increasing the interaction strength. The difference between the two cases is shown to originate from the momentum dependence of the $p$-wave interaction, which vanishes in the low momentum limit. We also identify the pseudogap regime in the phase diagram with respect to the temperature and the $p$-wave interaction strength. Since the pseudogap is a precursor phenomenon of the superfluid phase transition, our results would be useful for the research toward the realization of $p$-wave superfluid Fermi gases. 
\end{abstract}
\maketitle
\section{Introduction}
Since the discovery of $p$-wave Feshbach resonances in $^{40}$K\cite{Regal,Ticknor} and $^6$Li\cite{Zhang,Schunck} Fermi gases, the possibility of $p$-wave superfluid state has been extensively discussed in cold atom physics\cite{Ohashi,Ho,Gurarie,Gurarie2,Levinsen,Botelho,Iskin,Iskin2,Iskin3,Grosfeld,Mizushima,Mizushima2,Mizushima3,Han,Cheng,Cheng2,Maier,Gunter}. Once the $p$-wave superfluid phase is realized in this system, one can examine various superfluid properties of this unconventional pairing state from the weak-coupling regime to the strong-coupling limit in a unified manner, by adjusting the threshold energy of a $p$-wave Feshbach resonance. The existence of various $p$-wave superfluid phases makes us expect much richer physics than the case of the isotropic $s$-wave superfluid. Since unconventional Cooper pairings are important issues in metallic superconductivity, as well as in superfluid $^3$He, the realization of a highly tunable $p$-wave superfluid would make a great impact on, not only cold atom physics, but also condensed matter physics. So far, the $p$-wave superfluid Fermi gas has not been realized yet. However, $p$-wave molecules have been recently observed in $^{40}$K\cite{Regal2,Geabler} and $^{6}$Li\cite{Zhang,Inaba,Fuchs} Fermi gases.
\par
In the current stage of research, one of the most important issues is to reach the $p$-wave superfluid phase transition temperature $T_{\rm c}$. In this regard, in order to see to what extent the current experimental situation is close to this goal, the observation of a precursor phenomenon of the $p$-wave superfluid above $T_{\rm c}$ would be helpful. That is, when one increases the strength of the $p$-wave pairing interaction associated with a $p$-wave Feshbach resonance, strong pairing fluctuations are expected to cause the pseudogap phenomenon, where a superfluid-gap like structure appears in single-particle excitation spectra even in the normal state. Since the pseudogap temperature $T^*$, which is defined as the temperature below which the pseudogap appears, is higher than $T_{\rm c}$, the former would be experimentally more accessible than the latter. In addition, since the pseudogap becomes more remarkable near $T_{\rm c}$, the detailed observation of the pseudogap enables us to estimate how the current experiment is close to $T_{\rm c}$. Thus, besides the importance as a typical strong-coupling phenomenon, the pseudogap would be also important for the research toward the realization of $p$-wave superfluid Fermi gases. 
\par
The pseudogap has been recently discussed in the BCS (Bardeen-Cooper-Schrieffer)-BEC (Bose-Einstein condensation) crossover regime of an ultracold Fermi gas with an $s$-wave interaction\cite{Stewart,Gaebler,Perali2,Salomon,Tsuchiya1,Tsuchiya2,Tsuchiya3,Levin,Hui,Bulgac}. Although the existence of the pseudogap in this system is still in debate\cite{Salomon}, it has been pointed out that the anomalous single-particle excitation spectra observed in the crossover region\cite{Stewart,Gaebler,Perali2} may be explained as a pseudogap phenomenon, originating from strong pairing fluctuations\cite{Perali2,Tsuchiya1,Tsuchiya2,Hui,Levin,Bulgac}.
\par
\begin{figure}
\centerline{\includegraphics[width=10cm]{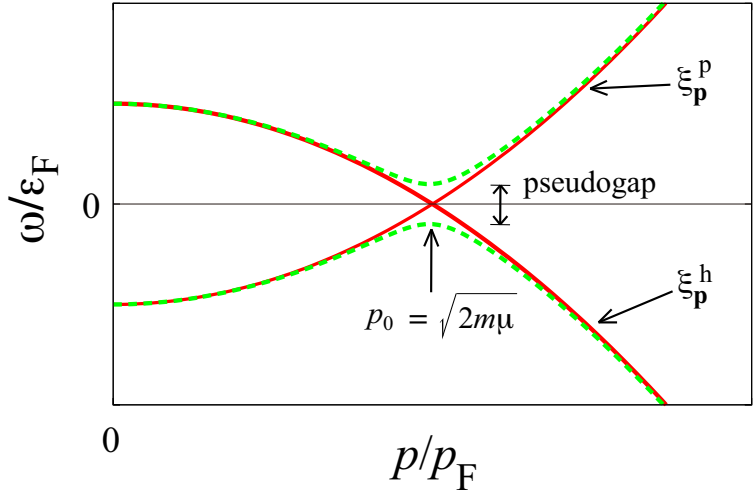}}
\caption{(Color online) Schematic explanation for the pseudogap phenomenon. Pairing fluctuations couple the particle excitation spectrum $\xi_{\bf p}^{\rm p}={\bf p}^2/(2m)-\mu$ with a hole excitation spectrum $\xi^{\rm h}_{\bf p}=-[{\bf p}^2/(2m)-\mu]$ around ${\tilde p}_{\rm F}=\sqrt{2m\mu}$ (when $\mu\ge 0$). This induces a gap-like structure in single-particle excitations around $\omega=0$ (two dashed lines). Since pairing fluctuations also induce a finite lifetime of single-particle states, finite intensity actually exists inside this gap, giving a pseudogap structure.}
\label{Fig1}
\end{figure}

Because of the momentum dependence of the $p$-wave interaction, we can expect that the $p$-wave pseudogap phenomenon is very different from the $s$-wave case. To briefly see this, we first note that the pseudogap phenomenon can be physically understood as a result of a particle-hole coupling induced by pairing fluctuations\cite{Tsuchiya3}. That is, as schematically shown in Fig.\ref{Fig1}, pairing fluctuations couple particle excitations ($\xi_{\bf p}^{\rm p}={\bf p}^2/(2m)-\mu$) with hole excitations ($\xi^{\rm h}_{\bf p}=-[{\bf p}^2/(2m)-\mu]$) around the momentum ${\tilde p}_{\rm F}\equiv\sqrt{2m\mu}$, at which the two branches cross with each other (where $m$ is a particle mass, and $\mu$ is the Fermi chemical potential). This coupling effect leads to level repulsion around ${\tilde p}_{\rm F}$, leading to a gap-like structure at $\omega\simeq 0$ in the single-particle excitation spectrum (dashed lines in Fig.\ref{Fig1}). In the ordinary $s$-wave case with a constant interaction $-U$, the particle-hole coupling is only dominated by the magnitude of $U$, so that the pseudogap simply becomes larger for a stronger $U$. On the other hand, in the $p$-wave case, since the interaction has the form $-U{\bf p}\cdot{\bf p}'$\cite{Ohashi,Ho}, the interaction strength around ${\tilde p}_{\rm F}$ is evaluated as $\sim U{\tilde p}_{\rm F}^2=2mU\mu$ (when $\mu>0$). Then, while the increase of $U$ promotes the pseudogap phenomenon, the decrease of $\mu$ by the strong-coupling effect\cite{Ohashi,Ho} suppresses the particle-hole coupling. Thus, the $p$-wave pseudogap phenomenon is expected to involve the competition between the increase of $U$ and the decrease of $\mu$.
\par
In this paper, we theoretically investigate the pseudogap phenomenon in a one-component Fermi gas with a tunable $p$-wave interaction. Extending the strong-coupling $T$-matrix theory for the $s$-wave interaction\cite{Tsuchiya1,Tsuchiya2,Tsuchiya3,Perali} to the $p$-wave case, we calculate the single-particle density of states, as well as the spectral weight, in the normal state above $T_{\rm c}$.  We show how the pseudogap develops in these quantities near $T_{\rm c}$, as one increases the interaction strength. From the temperature dependence of the pseudogap, we determine the pseudogap temperature $T^*$, and identify the pseudogap region in the phase diagram with respect to the temperature and the $p$-wave interaction strength.
\par
The outline of this paper is as follows. In Sec.II, we explain our strong-coupling $T$-matrix theory for a one-component uniform Fermi gas with a $p$-wave interaction. In Sec. III, we examine the pseudogap seen in the single-particle density of states. We also determine the pseudogap temperature $T^*$ to identify the pseudogap regime in the phase diagram of a $p$-wave Fermi gas. In Sec.IV, we consider the single-particle spectral weight. Throughout this paper, we set $\hbar=k_{\rm B}=1$, and the system volume $V=1$.
\par
\par
\section{Formulation}
We consider a one-component uniform Fermi gas with a $p$-wave pairing interaction, described by the Hamiltonian
\begin{equation}
H=\sum_{\bf p} \xi_{\bf p}c_{\bf p}^{\dagger}c_{\bf p}
-\frac{U}{2}\sum_{{\bf p},{\bf p'},{\bf q}}{\bf p}\cdot{\bf p}'
c_{\bf p + \frac{\bf q}{2}}^{\dagger}c_{-\bf p + \frac{\bf q}{2}}^{\dagger}
c_{\bf p' + \frac{\bf q}{2}}c_{-\bf p' + \frac{\bf q}{2}}.
\label{eq.1}
\end{equation}
Here, $c^\dagger_{\bf p}$ is the creation operator of a Fermi atom with the kinetic energy $\xi_{\bf p}=\varepsilon_{\bf p}-\mu=p^2/2m-\mu$, measured from the Fermi chemical potential $\mu$ (where $m$ is an atomic mass). $-U{\bf p \cdot p'}$ ($U > 0$) is a $p$-wave attractive interaction. In this paper, we treat $U$ as a tunable parameter, by implicitly assuming that this  interaction is associated with a $p$-wave Feshbach resonance. More generally, the $p$-wave interaction may be written as $-\sum_{j=x,y,z}U_jp_jp_j'$. Experimentally, the splitting of a $p$-wave Feshbach resonance by a magnetic dipole-dipole interaction has been observed\cite{Ticknor}, indicating that all of the three components $U_{j=x,y,z}$ do not have the same magnitude. However, we ignore this anisotropy, for simplicity, and set $U_x=U_y=U_z=U$ in this paper.
\par
Since the $p$-wave interaction in Eq. (\ref{eq.1}) involves an ultraviolet divergence, it is convenient to measure the interaction strength in terms of the scattering volume $v$ and the effective range $k_0$\cite{Ticknor}. These are related to the coupling $U$ as
\begin{eqnarray}
\frac{4\pi v}{m}&=&-\frac{U}{3-U\sum_{\bf p}^{p_{\rm c}} \frac{p^2}{2\varepsilon_{\bf p}}},
\label{eq.2_1}
\\
k_0&=&-\frac{4\pi}{m^2}\sum_{\bf p}^{p_{\rm c}} \frac{p^2}{2\varepsilon_{\bf p}^2},
\label{eq.2_2}
\end{eqnarray}
where $p_{\rm c}$ is a momentum cutoff. Following the experimental result on a $^{40}$K Fermi gas\cite{Ticknor}, we take $k_0/p_{\rm F}=-30$, where $p_{\rm F}$ is the Fermi momentum. As usual, the interaction strength is conveniently measured in term of  $1/(vp_{\rm F}^3)$\cite{Ohashi,Ho}. The increase of this quantity corresponds to the increase of $U$.
\par 
\begin{figure}
\centerline{\includegraphics[width=10cm]{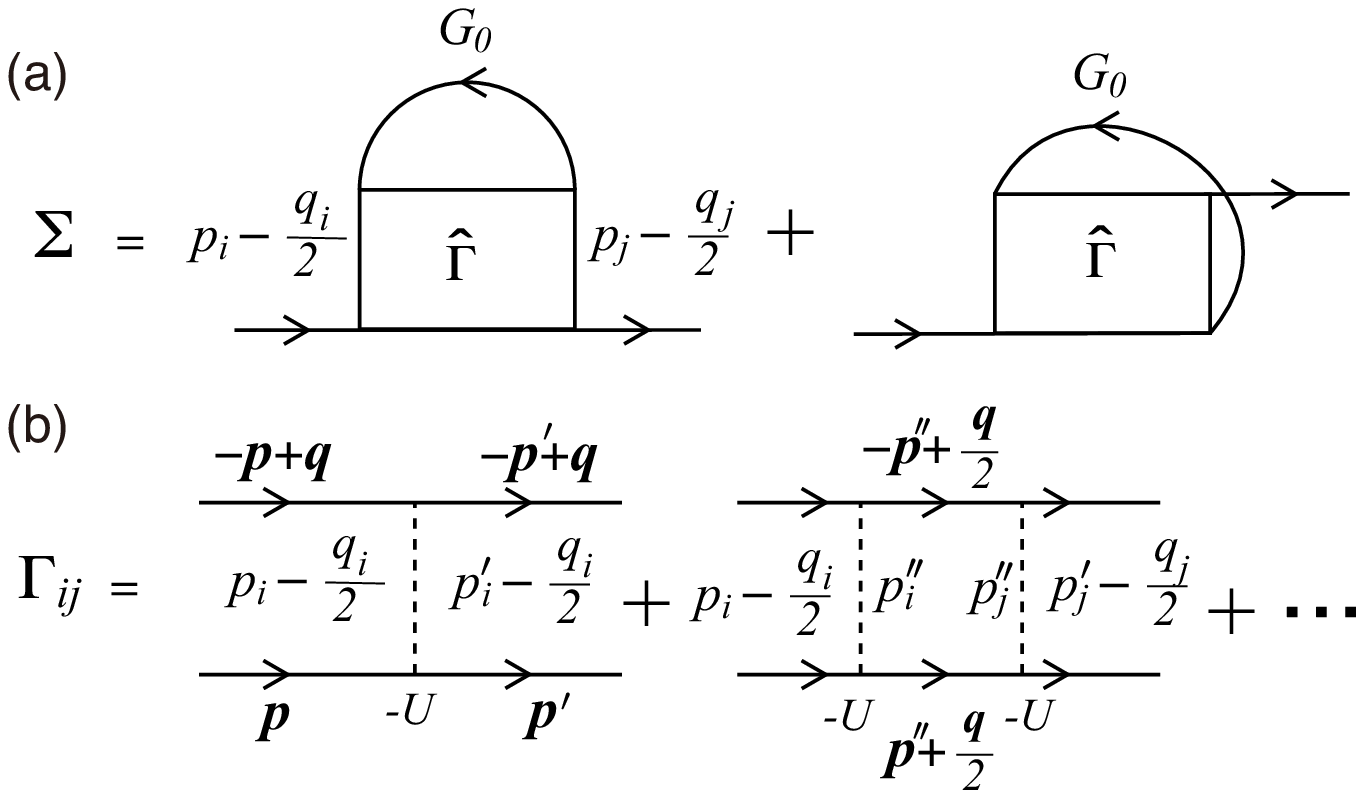}}
\caption{(a) Self-energy $\Sigma({\bf p},i\omega_m)$, and (b) particle-particle scattering matrix $\Gamma({\bf q},i\nu_n)$ in the $T$-matrix theory. The solid line and the dashed line describe the single-particle Green's function $G_0$ and the $p$-wave interaction $-U{\bf p \cdot p'}$, respectively.}
\label{fig2}
\end{figure}
\par
We treat the $p$-wave interaction within the $T$-matrix theory. Although the validity of this theory for the $p$-wave system is still unclear, at least in the $s$-wave case, it has been shown that it correctly describes the pseudogap phenomenon. In particular, the calculated single-particle excitation spectrum well agrees with the recent photoemission-type experiment on $^{40}$K Fermi gases\cite{Perali2,Tsuchiya1,Tsuchiya2,Tsuchiya3}. The single-particle thermal Green's function is given by
\begin{equation}
G({\bf p},i\omega_m)=\frac{1}{i\omega_m-\xi_{\bf p}-\Sigma({\bf p},i\omega_m)},
\label{eq.3}
\end{equation}
where $\omega_m$ is the fermion Matsubara frequency. The self-energy correction $\Sigma({\bf p},i\omega_m)$ in the $T$-matrix theory is diagrammatically given in Fig.\ref{fig2}. Summing up these diagrams, we obtain 
\begin{eqnarray}
\Sigma({\bf p},i\omega_m)&=&\frac{2}{\beta} \sum_{{\bf q},i\nu_n}\sum_{i,j=x,y,z} \left[
\left(p_i- \frac{q_i}{2}  \right) 
\Gamma_{ij}({\bf q},i\nu_n)
\left(p_j- \frac{q_j}{2}  \right) 
\right]
\nonumber
\\
&\times&
G_0({\bf -p+q},-i\omega_m+i\nu_n),
\label{eq.4}
\end{eqnarray}
where $\beta=1/T$, and $\nu_n$ is the boson Matsubara frequency. $G^0=(i\omega_m - \xi_{\bf p})^{-1}$ is the Green's function for a free Fermi gas. The $3\times 3$-matrix particle-particle scattering matrix ${\hat \Gamma}=\{\Gamma\}_{ij}$ is given by
\begin{eqnarray}
\hat{\Gamma}({\bf q},i\nu_n)&=&-\frac{U}
{1-U\hat{\Pi}({\bf q},i\nu_n)}.
\label{eq.5}
\end{eqnarray}
Here, the $3\times 3$-matrix correlation function ${\hat \Pi}=\{\Pi\}_{ij}$ describes fluctuations in the $p$-wave Cooper channel, given by
\begin{equation}
\Pi_{ij}({\bf q},i\nu_n)=\sum_{\bf k}k_ik_j
\frac{1-f(\xi_{\bf k + q/2})-f(\xi_{\bf -k+q/2})}
{\xi_{\bf k+q/2}+\xi_{\bf -k+q/2}+i\nu_n},
\label{eq.6}
\end{equation}
where $f(\xi_{\bf k})$ is the Fermi distribution function.
\par
\begin{figure}
\centerline{\includegraphics[width=8cm]{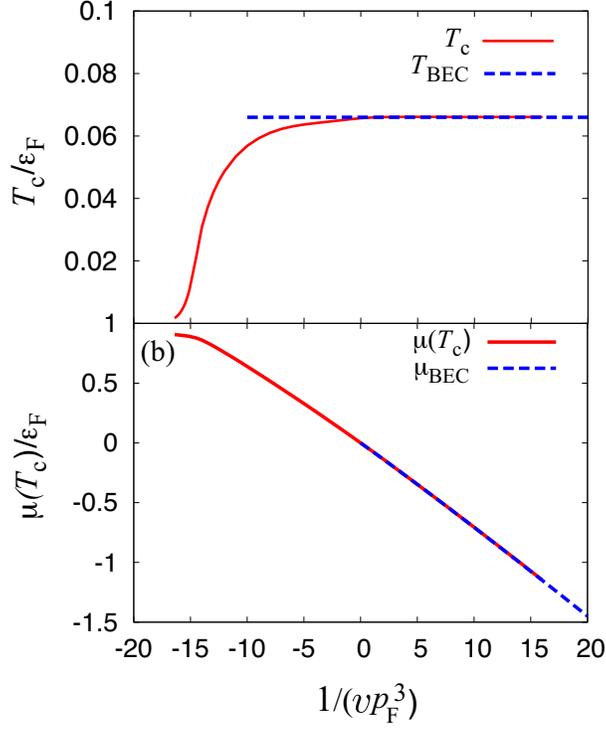}}
\caption{(Color online) Calculated $T_{\rm c}$ (a), and $\mu(T_{\rm c})$ (b), as functions of the interaction strength, measured in term of the scattering volume $v$. We set the effective range $k_0=-30p_{\rm F}$\cite{Ticknor}. $T_{\rm BEC}$ and $\mu_{\rm BEC}$ are given by Eqs. (\ref{eq.BEC}) and (\ref{eq.12}), respectively.}
\label{fig3}
\end{figure}

To examine the pseudogap in the normal state, we first determine the superfluid phase transition temperature $T_{\rm c}$. The equation for $T_{\rm c}$ is obtained from the Thouless criterion, stating that the particle-particle scattering matrix ${\hat \Gamma}$ at $T_{\rm c}$ has a pole at ${\bf q}=\nu_n=0$. Using that ${\hat \Pi}$ is diagonal at ${\bf q}=\nu_n=0$, one obtains the $T_{\rm c}$-equation as
\begin{equation}
1=\frac{U}{3} \sum_{\bf p}^{p_c}  
\frac{p^2}{2\xi_{\bf p}}
\tanh \frac{\beta \xi_{\bf p}}{2} 
.
\label{eq.10}
\end{equation}
\par
As shown in Refs.\cite{Ohashi,Ho}, the Fermi chemical potential $\mu$ remarkably deviates from the Fermi energy $\varepsilon_{\rm F}$, as one goes away from the weak-coupling regime. This strong coupling effect is conveniently taken into account by solving Eq. (\ref{eq.10}), together with the equation for the number $N$ of Fermi atoms, 
\begin{equation}
N={1 \over \beta}
\sum_{{\bf p},i\omega_m} G({\bf p},i\omega_m) e^{i\omega_m\delta}.
\label{eq.11}
\end{equation}
\par
Figure \ref{fig3} shows the self-consistent solutions for $T_{\rm c}$ and $\mu(T_{\rm c})$, calculated from the coupled equations (\ref{eq.10}) and (\ref{eq.11}). In panel (a), while $T_{\rm c}$ increases with increasing the interaction strength in the weak coupling regime ($1/(vp_{\rm F}^3)\ll-1$), it approaches a constant value in the strong coupling regime ($1/(vp_{\rm F}^3)\gesim 0$). In the strong-coupling limit ($1/(vp_{\rm F}^3)\to\infty$), the system can be regarded as a Bose gas, consisting of three kinds of tightly bound molecules formed by the pairing interactions $-Up_jp_j'$ ($j=x,y,z$)\cite{note1}. The number $N_{\rm B}$ of molecules in each component equals $N_{\rm B}=N/6$, so that $T_{\rm c}$ in this limit ($\equiv T_{\rm BEC}$) is found to be
\begin{equation}
T_{\rm BEC}={2\pi \over \zeta(3/2)}{N_{\rm B}^{2/3} \over M}=0.066\varepsilon_{\rm F},
\label{eq.BEC}
\end{equation}
where $M=2m$ is a molecular mass, and $\zeta(3/2)=2.612$ is the zeta-function. Indeed, Fig.\ref{fig3}(a) shows that $T_{\rm c}\simeq T_{\rm BEC}$ when $1/(vp_{\rm F}^3)\gesim 0$.
\par
Such a molecular formation can be also seen in Fig.\ref{fig3}(b). In this panel, the chemical potential $\mu$ decreases to be negative in the strong-coupling regime. When the pairing interaction is very strong ($1/(vp_{\rm F}^3)\gg 1$), Eqs.(\ref{eq.10}) and (\ref{eq.11}) give
\begin{equation}
\mu= -\frac{1}{mv|k_0|} \left[
1+2\sqrt{\frac{2}{|k_0|^3 v}}+ O\left(\frac{1}{|k_0|^3 v}
\right)^{\frac{3}{2}} 
\right].
\label{eq.12}
\end{equation}
As expected, the limiting value $|\mu|\to1/(mv|k_0|)$ equals half the binding energy of a {\it two-body} $p$-wave bound state. Figure \ref{fig3}(b) shows that $\mu(T_{\rm c})$ is well described by Eq. (\ref{eq.12}), when $1/(vp_{\rm F}^3)\gesim 0$.
\par
\begin{figure}
\centerline{\includegraphics[width=10cm]{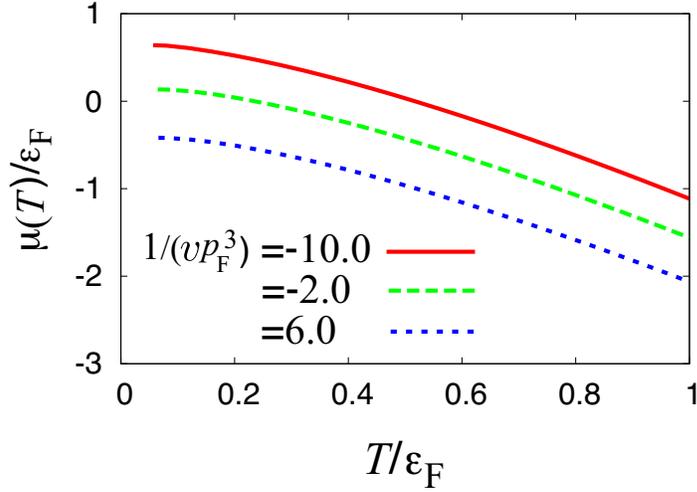}}
\caption{(Color online) Calculated Fermi chemical potential $\mu(T\ge T_{\rm c})$ as a function of temperature.}
\label{fig4}
\end{figure}

Once $T_{\rm c}$ is determined, we calculate the chemical potential above $T_{\rm c}$, by only solving the number equation (\ref{eq.11}). We show the calculated $\mu(T\ge T_{\rm c})$ in Fig.\ref{fig4}. The single-particle density of states (DOS) $\rho(\omega)$, as well as the single-particle spectral weight (SW) $A({\bf p},\omega)$, are then calculated from the analytic continued Green's function as, respectively,
\begin{eqnarray}
\rho(\omega)&=&-\frac{1}{\pi}\sum_{\bf p}{\rm Im} [G({\bf p},i\omega_m \to \omega + i\delta)],
\label{eq.13_1}
\\
A({\bf p}, \omega)&=&-\frac{1}{\pi}{\rm Im} [G({\bf p},i\omega_m \to \omega + i\delta)].
\label{eq.13_2}
\end{eqnarray}
\par
The $p$-wave superfluid order parameter is known to be anisotropic in  momentum space, reflecting the momentum-dependent pairing interaction $-U{\bf p}\cdot{\bf p}'$\cite{Vollhardt}. Thus, single-particle excitations below $T_{\rm c}$ are also anisotropic in momentum space (except for some special pairing states, such as the Balian-Werthamer (BW) phase). However, we point out that single-particle excitations above $T_{\rm c}$ are {\it isotropic} even in the presence of the $p$-wave interaction $-U{\bf p}\cdot{\bf p}$. To show this, we note that the pair-correlation function $\Pi_{ij}$ in Eq. (\ref{eq.6}) becomes diagonal, when we choose the $z$-axis along the ${\bf q}$-direction. Under this choice, the particle-particle scattering matrix $\Gamma_{ij}$ in (\ref{eq.5}) becomes diagonal, where the diagonal components are given by
\begin{eqnarray}
\Gamma_{ii}({\bf q},i\nu_n)&=&\frac{1}
{\frac{m}{12\pi v}
+\Pi_{ii}(q,i\nu_n)
+\frac{m}{18\pi^2}\left(\frac{\pi k_0}{4} \right)^3},
\label{eq.9a}
\end{eqnarray}
where the interaction is described by the scattering volume $v$ and the effective range $k_0$. In Eq. (\ref{eq.9a}), the diagonal correlation function $\Pi_{ii}$ is given by 
\begin{eqnarray}
\Pi_{ii}({\bf q},i\nu_n)&=&\sum_{\bf k}k_i^2
\frac{1-f(\Xi_+)-f(\Xi_-)}
{2\varepsilon_k + \frac{\varepsilon_q}{2}-2\mu+i\nu_n}.
\label{eq.9b}
\end{eqnarray}
In Eq. (\ref{eq.9b}), $\Xi_{\pm}=[k^2 \pm kq\cos \theta_{kq} +q^2/4]/(2m)-\mu$, where $\theta_{kq}$ is the angle between ${\bf k}$ and ${\bf q}$. Since Eq. (\ref{eq.9a}) does not depend on the direction of ${\bf q}$, one may simply write $\Gamma_\perp(q,i\nu_n)\equiv\Gamma_{xx}({\bf q},i\nu_n)=\Gamma_{yy}({\bf q},i\nu_n)$ and $\Gamma_\parallel(q,i\nu_n)\equiv\Gamma_{zz}({\bf q},i\nu_n)$. Substituting these expressions into Eq. (\ref{eq.4}), we obtain
\begin{eqnarray}
\Sigma({\bf p},i\omega_m)=\frac{2}{\beta}\sum_{{\bf q}, i\nu_n} 
\left[
p^2\sin^2\theta_{pq} \Gamma_\perp(q,i\nu_n)
+\left(
p\cos\theta_{pq}-\frac{q}{2}
\right)^2 \Gamma_\parallel(q,i\nu_n)
\right]&
\nonumber
\\
\times
\frac{1}{\displaystyle -i\omega_m +i\nu_n - \frac{1}{2m}
\left( p^2-2pq\cos \theta_{pq} + q^2 \right) + \mu},
\label{eq.7}
\end{eqnarray}
where $\theta_{pq}$ is the angle between ${\bf p}$ and ${\bf q}$. Executing the angular integration with respect to $\theta_{pq}$ in Eq. (\ref{eq.7}), one finds that the self-energy $\Sigma({\bf p},i\omega_m)$ is {\it isotropic} in momentum space. Thus, the spectral weight $A({\bf p},\omega)$ in Eq. (\ref{eq.13_2}) is also isotropic in momentum space. When all of the coupling constants $U_j$ $(j=x,y,z)$ do not have the same value, the spectral weight becomes anisotropic. 

\par
\begin{figure}
\centerline{\includegraphics[width=10cm]{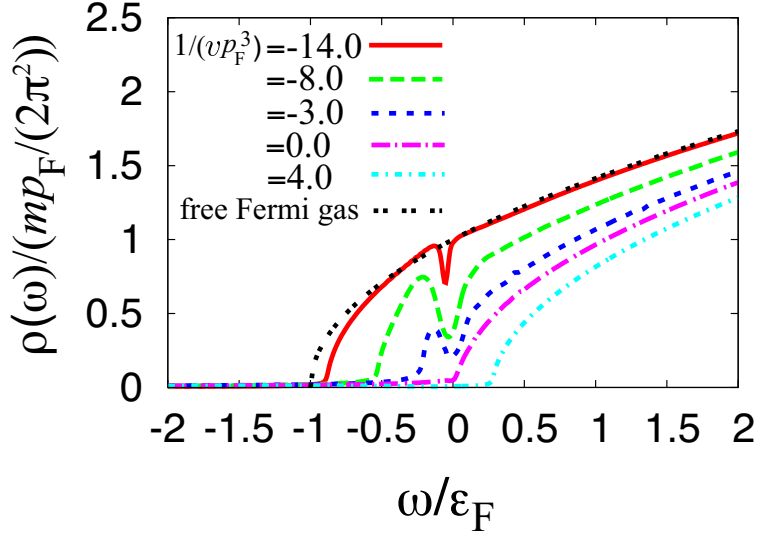}}
\caption{(Color online)  Calculated single-particle density of states (DOS) $\rho(\omega)$ at $T_{\rm c}$. In this figure, `free Fermi gas' shows DOS of a non-interacting Fermi gas with $\mu=\varepsilon_{\rm F}$.}
\label{fig5}
\end{figure}

\section{density of states and $p$-wave pseudogap phenomenon}

Figure \ref{fig5} shows the single-particle density of states (DOS) $\rho(\omega)$ at $T_{\rm c}$. In the weak-coupling case ($1/(vp_{\rm F}^3)=-14.0$), we see a small dip structure around the $\omega=0$. Since the superfluid order parameter vanishes at $T_{\rm c}$, this dip structure is just the pseudogap associated with $p$-wave pairing fluctuations. This pseudogap becomes more remarkable when $1/(vp_{\rm F}^3)=-8.0$. However, as one further increases the interaction strength ($1/(vp_{\rm F}^3)=-3$), the pseudogap becomes obscure. When $1/(vp_{\rm F}^3)\gesim 0$, apart from the weak intensity in the negative energy region, the overall structure is similar to DOS for a free Fermi gas with a negative chemical potential, $\rho(\omega)\propto\sqrt{\omega+|\mu|}\Theta(\omega-|\mu|)$ (where $\Theta(x)$ is the step function)\cite{note2}. 
\par
We emphasize that the present non-monotonic interaction dependence of the pseudogap phenomenon is quite different from the ordinary $s$-wave case. In the latter, the pseudogap simply develops, as one passes through the BCS-BEC crossover region\cite{Tsuchiya1,Tsuchiya2,Tsuchiya3,Perali2}.  
\par
As briefly discussed in Sec.I, the momentum dependence of the $p$-wave interaction is the key to understand the difference between the $s$-wave case and the $p$-wave case. To see this more clearly, it is convenient to approximately treat the self-energy in Eq. (\ref{eq.7}) as, using the fact that the particle-particle scattering matrix $\Gamma_{ii}({\bf q}=0,i\nu_n=0)$ in Eq. (\ref{eq.9a}) diverges at $T_{\rm c}$ (Thouless criterion), 
\begin{eqnarray}
\Sigma({\bf p},i\omega_m)&\simeq&
-\Delta_{\rm pg}^2({\bf p})G_0({\bf -p},-i\omega_m).
\label{eq.pseudo}
\end{eqnarray}
Here, $\Delta^2_{\rm pg}({\bf p})\equiv p^2{\tilde \Delta}^2_{\rm pg}$ is the so-called pseudogap parameter\cite{Perali}, where
\begin{equation}
\tilde{\Delta}_{\rm pg}^2=- \frac{1}{\beta}\sum_{{\bf q}, i\nu_n} 
\left[
\Gamma_\perp(q,i\nu_n)
+\Gamma_\parallel(q,i\nu_n)
\right].
\label{eq.3.2}
\end{equation}
Substituting Eq. (\ref{eq.pseudo}) into Eq. (\ref{eq.3}), one finds
\begin{equation}
G({\bf p},i\omega_m)=
{1 \over \displaystyle (i\omega_m-\xi_{\bf p})-
{\Delta_{\rm pg}^2({\bf p}) \over i\omega_m+\xi_{\bf p}}}.
\label{eq.pseudo2}
\end{equation}
In Eq. (\ref{eq.pseudo2}), $1/(i\omega_m-\xi_{\bf p})$ and $1/(i\omega_m+\xi_{\bf p})$ are just the particle and hole Green's functions, respectively. From Eq. (\ref{eq.pseudo2}), pairing fluctuations are found to couple the particle branch $\omega=\xi_{\bf p}$ with the hole branch $\omega=-\xi_{\bf p}$ around $\omega=0$ (when $\mu>0$), with the coupling strength $\Delta_{\rm pg}^2({\bf p})$. This coupling effect naturally leads to the level repulsion between the two branches. Indeed, Eq. (\ref{eq.pseudo2}) gives the BCS-type gapped excitations,
\begin{equation}
E_{\bf p}^\pm=\pm\sqrt{\xi_{\bf p}^2+\Delta_{\rm pg}^2({\bf p})}.
\label{eq.pseudo3}
\end{equation}
Defining $\Delta E$ as the minimum value of the energy gap between $E_{\bf p}^+$ and $E_{\bf p}^-$, one finds
\begin{eqnarray}
\Delta E =
\left\{
\begin{array}{ll}
2{\tilde \Delta}_{\rm pg}\sqrt{2m\mu-m^2\tilde{\Delta}_{pg}^2}& ~~~~~(\mu > m{\tilde \Delta}_{\rm pg}), \\
2\left| \mu \right|& ~~~~~(\mu < m{\tilde \Delta}_{\rm pg}).
\end{array}
\right.
\label{eq.3.4}
\end{eqnarray}
We briefly note that pairing fluctuations actually induce a finite lifetime of quasi-particle excitations, which broadens excitation spectra (although this effect is ignored in the simple approximation in Eq. (\ref{eq.pseudo})). The resulting DOS has a finite intensity inside the gap $\Delta E$ in Eq. (\ref{eq.3.4}), so that the {\it pseudogap} is realized. 
\par
Since pairing fluctuations are weak in the weak-coupling regime, one may safely take $\mu\simeq\varepsilon_{\rm F}\gg \Delta_{\rm pg}$ in Eq. (\ref{eq.3.4}). In this case, one finds $\Delta E\simeq 2p_{\rm F} \tilde{\Delta}_{\rm pg}=2\Delta_{\rm pg}(p=p_{\rm F})$. That is, the pseudogap is dominated by pairing fluctuations near the Fermi surface in this regime. 
\par
As one approaches the strong-coupling regime, while the strong pairing fluctuations enhances ${\tilde \Delta}_{\rm pg}$, they also decrease the magnitude of $\mu$, as shown in Fig.\ref{fig3}(b). Thus, the pseudogap phenomenon is dominated by the competition between the increase of ${\tilde \Delta}_{\rm pg}$ and the decrease of $\mu$, which leads to the non-monotonic behavior of the pseudogap structure shown in Fig.\ref{fig5}. 
\par
In the strong-coupling regime where the chemical potential is given by $\mu\simeq -1/(mvk_0)$ (See Eq. (\ref{eq.12}).), $\Delta E=2|\mu|=2/(mvk_0)$ just equals the binding energy of a two-body $p$-wave bound state. That is, the physical meaning of $\Delta E$ continuously changes from the pseudogap size to the binding energy of a two-body bound molecule with increasing the interaction strength.
\par
We note that, when $\mu<m{\tilde \Delta}_{\rm pg}$, the minimum gap energy $\Delta E=2|\mu|$ in Eq. (\ref{eq.3.4}) is obtained at ${\bf p}=0$. Considering the region around ${\bf p}=0$, one may ignore $\Delta_{\rm pg}({\bf p}\sim 0)$ in Eq. (\ref{eq.pseudo2}). This explains why DOS shown in Fig.\ref{fig5} is similar to DOS for a free Fermi gas when $1/(vp_{\rm F}^3)\gesim 0$ (where $\mu<0$ is realized, as shown in Fig.\ref{fig3}(b)).
\par
Here, we briefly compare the present result with the case of an $s$-wave interaction. The above discussion is also applicable to the $s$-wave case, where the pseudogap parameter $\Delta_{\rm pg}(p)$ is replaced by the ${\bf p}$-independent expression,
\begin{equation}
\Delta_{\rm pg}^2=-T\sum_{{\bf q},i\nu_n}\Gamma({\bf q},i\nu_n),
\label{eq.s}
\end{equation}
where $\Gamma({\bf q},i\nu_n)$ is the $s$-wave particle-particle scattering matrix\cite{Tsuchiya3,Perali3}. In this case, the pseudogap width is simply evaluated as 2$\Delta_{\rm pg}$ when $\mu>0$, which {\it monotonically} increases with increasing the interaction strength. When $\mu<0$ in the strong-coupling regime, one finds $\Delta E=2\sqrt{|\mu|^2+\Delta_{\rm pg}^2}$. In the strong-coupling BEC limit, since $|\mu|$ becomes much larger than $\Delta_{\rm pg}$, the gap size eventually reduces to the binding energy of a two-body bound states $E_g=2|\mu|=1/(ma_s^2)$\cite{Tsuchiya3,Perali3}, where $a_s$ is the $s$-wave scattering length.
\par

\begin{figure}
\centerline{\includegraphics[width=8cm]{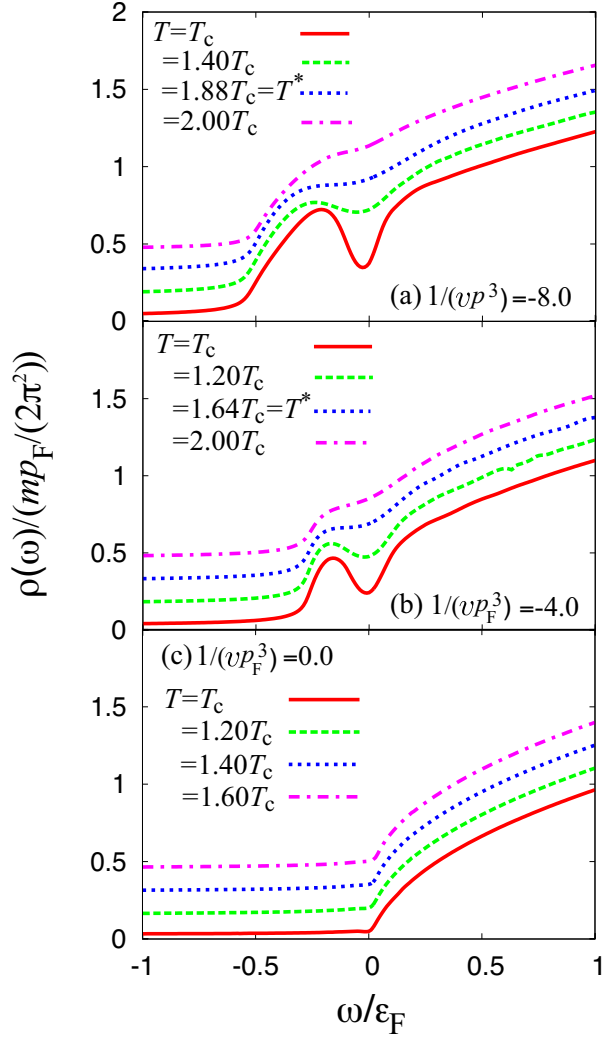}}
\caption{(Color online) Single-particle density of states (DOS) $\rho(\omega)$ above $T_{\rm c}$. In each panel, $T_{\rm c}/\varepsilon_{\rm F}$ is given by (a) 0.061, (b) 0.064 , and (c) 0.066. $T^*$ is the pseudogap temperature.}
\label{fig6}
\end{figure}

\begin{figure}
\centerline{\includegraphics[width=8cm]{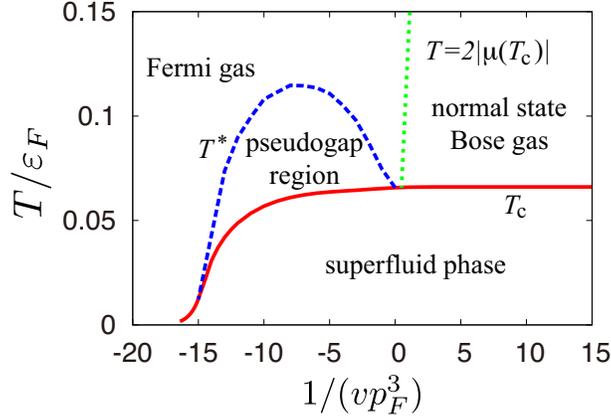}}
\caption{(Color online) Phase diagram of a one-component Fermi gas with a $p$-wave pairing interaction. The pseudogap region is surrounded by the pseudogap temperature $T^*$ and the superfluid phase transition temperature $T_{\rm c}$. In this figure, we also draw the line along $2|\mu(T_{\rm c})|$ above $T_{\rm c}$ when $\mu<0$. Physically, this gives a characteristic temperature where two-body bound molecules are formed. The right side of this line may be thus regarded as a normal state molecular Bose gas, rather than a Fermi gas. In this figure, while $T_{\rm c}$ is a phase transition temperature, $T^*$ and $2|\mu|$ are crossover temperatures, without being accompanied by any phase transition.}
\label{fig7}
\end{figure}
 
\par
Since pairing fluctuations become weak with increasing the temperature above $T_{\rm c}$, the pseudogap structure in DOS gradually becomes obscure, as shown in Fig.\ref{fig6}. When we define the pseudogap temperature $T^*$ as the temperature at which the dip structure in DOS disappears, we can identify the pseudogap region in the phase diagram with respect to the temperature and the interaction strength, as shown in Fig.\ref{fig7}. As expected from the non-monotonic behavior of the pseudogap structure at $T_{\rm c}$, $T^*$ also exhibits non-monotonic interaction dependence. In the $s$-wave case, $T^*$ monotonically increases with increasing the interaction strength\cite{Tsuchiya1,Tsuchiya2,Tsuchiya3}.
\par
In Fig.\ref{fig7}, we also draw the line $T=2|\mu(T_{\rm c})|$ above $T_{\rm c}$ in the strong coupling regime where $\mu(T_{\rm c})<0$. Since $2|\mu|$ physically describes the binding energy of a two-body bound molecule in this regime, it gives a characteristic temperature below which two-body bound states are formed, overwhelming the thermal dissociation. Thus, the right side of this line may be regarded as a molecular Bose gas (which is not Bose condensed). We briefly note that $T=2|\mu|$, as well as $T^*$, are both crossover temperatures, without being accompanied by any phase transition. In Fig.\ref{fig7}, only $T_{\rm c}$ is the phase transition temperature.

\begin{figure}
\centerline{\includegraphics[width=7cm]{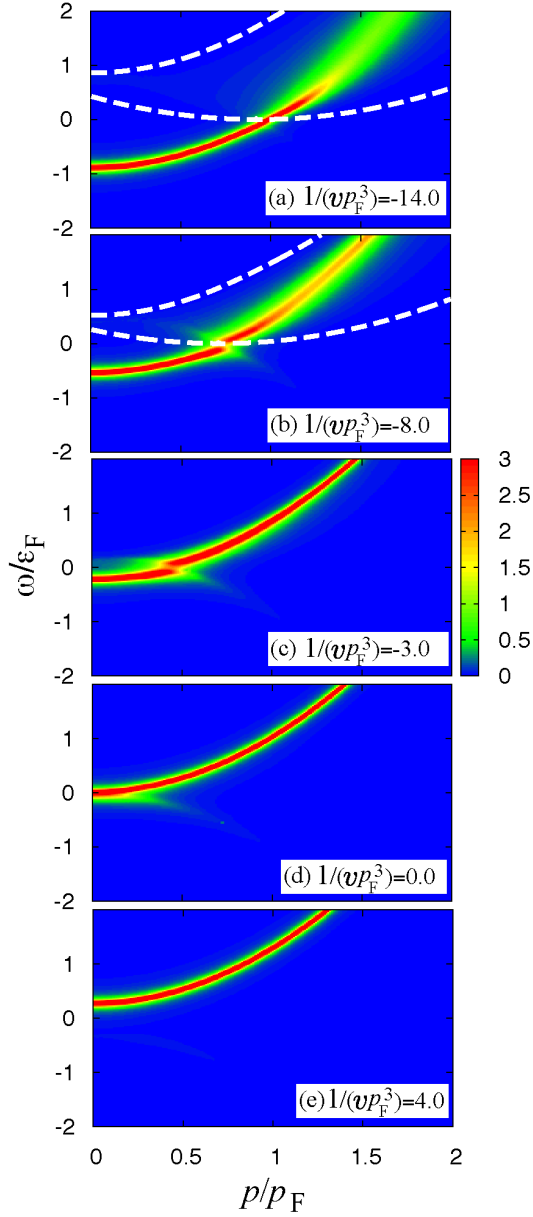}}
\caption{(Color online) Intensity of single-particle spectral weight $A({\bf p},\omega)$ at $T_{\rm c}$. In panels (a) and (b), the region between the two dashed lines satisfies Eq. (\ref{ap10}), where molecular excitations induce the broadening of the spectral peak.}
\label{fig8}
\end{figure}

\section{Spectral weight and coupling with molecular excitations}

Figure \ref{fig8} shows the single-particle spectral weight (SW) $A({\bf p},\omega)$ at $T_{\rm c}$. When $1/(vp_{\rm F}^3)=-14$ (panel (a)), while DOS already has a dip structure around $\omega=0$ (See Fig.\ref{fig5}.), such a pseudogap structure is still absent in SW. The spectral peak simply exists along the free-particle dispersion, $\omega=\varepsilon_{\bf p}-\mu$. 
\par
In panels (b) and (c), one sees a coupling between the particle excitations ($\omega=\varepsilon_{\bf p}-\mu$) and the hole excitations ($\omega=-[\varepsilon_{\bf p}-\mu]$), leading to the pseudogap around $\omega=0$. The momentum ${\tilde p}_{\rm F}$ at which the particle-hole coupling occurs is evaluated as ${\tilde p}_{\rm F}\simeq \sqrt{2m\mu}$, which is smaller for a stronger interaction, due to the decrease of $\mu$. When $\mu\simeq 0$ (panel (d)), because of the vanishing pseudogap parameter $\Delta_{\rm pg}(p)=p^2{\tilde \Delta}_{\rm pg}$ at ${\bf p}=0$, the pseudogap is no longer seen in SW. This result is consistent with the vanishing pseudogap in DOS at $1/(vp_{\rm F}^3)=0$, shown in Fig.\ref{fig5}.
\par
When one further increases the interaction strength (panel (e)), the peak line in the spectral weight is well fitted by the free-particle dispersion, $\omega=\varepsilon_{\bf p}+|\mu|$. Although the hole branch $\omega=-[\varepsilon_{\bf p}+|\mu|]$ also exists in the negative energy region, the intensity is much weaker than the particle branch. Indeed, using Eq. (\ref{eq.pseudo2}), one has
\begin{equation}
A({\bf p},\omega)=u_{\bf p}^2 \delta (\omega-E_{\bf p})+v_{\bf p}^2 \delta (\omega+E_{\bf p}),
\label{eq.3.3}
\end{equation}
where $E_{\bf p}=\sqrt{\xi_{\bf p}^2+p^2{\tilde \Delta}^2_{\rm pg}}$, $u_{\bf p}^2=[1+\xi_{\bf p}/E_{\bf p}]/2$, and $v_{\bf p}^2=[1-\xi_{\bf p}/E_{\bf p}]/2$. The first and second terms in Eq. (\ref{eq.3.3}) describe the contributions from particle excitations and hole excitations, respectively. Because $u_{\bf p}\gg v_{\bf p}$ in the strong-coupling regime (where $\mu/\varepsilon_{\rm F}\ll -1$), the first term in Eq. (\ref{eq.3.3}) gives dominant contribution. 
\par
\begin{figure}
\centerline{\includegraphics[width=8cm]{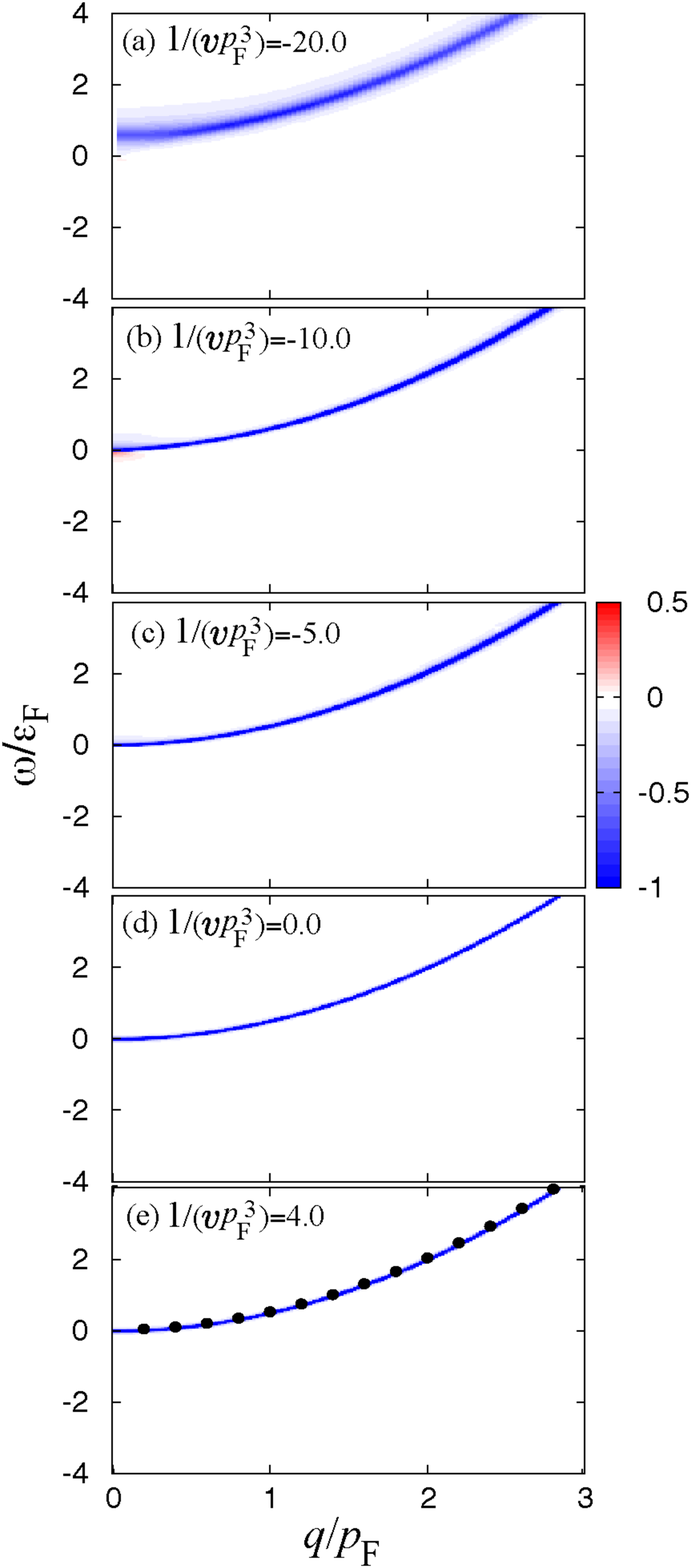}}
\caption{(Color online) Intensity of particle-particle scattering matrix ${\rm Im}[\Gamma_{\parallel}(q,i\nu_n\to\omega+i\delta)]$. Except for panel (a), we have used the analytic continued Eq. (\ref{eq.5}) at $T_{\rm c}$. For panel (a), since $T_{\rm c}$ is very small, it is very difficult to precisely calculate Eq. (\ref{eq.5}) at $T_{\rm c}$. To avoid this difficulty, we have used Eq. (\ref{ap.3}) in this case. In panel (e), solid circles show the dispersion of a free molecule, $\omega=q^2/(2M)$.
}
\label{fig9}
\end{figure}

Figure \ref{fig9} shows the spectra of the particle-particle scattering matrix, ${\rm Im}[\Gamma_\parallel(q,i\nu_n\to\omega+i\delta)]$ at $T_{\rm c}$, which physically describes molecular excitations. In each panel, one sees a sharp peak line, which is close to the dispersion of a free molecule, $\omega=q^2/(2M)$ with $M=2m$. Even in the weak-coupling regime shown in panel (a), the peak line is still close to the molecular dispersion, although the spectral peak is somehow broadened. 
\par
In the strong-coupling regime where tightly bound molecules have been already formed above $T_{\rm c}$, the sharp peak line along $\omega=q^2/(2M)$ is a reasonable result. Indeed, using $|\mu| \gg T_{\rm c}$ in Eq. (\ref{eq.9a}), we obtain
\begin{equation}
\Gamma_{\parallel}(q,i\nu_n) 
=\Gamma_{\perp}(q,i\nu_n)=
\frac{24\pi}{m^2 |k_0|}\frac{1}{\displaystyle i\nu_n-\left(\frac{q^2}{2M}-\mu_{\rm B} \right) }.
\label{eq.3.5}
\end{equation}
Equation (\ref{eq.3.5}) essentially has the same form as the single-particle Bose Green's function with a molecular mass $M=2m$ and the Bose chemical potential $\mu_{\rm B}= 2\mu-2/(mvk_0) \simeq 0$.
\par
To explain the molecular excitations seen in the weak-coupling regime, we recall that, our calculations are taking a relatively large value of the effective range $k_0$ ($|k_0|/p_{\rm F}=30\gg 1$), following the experimental result on $^{40}$K\cite{Ticknor}. Using this, and approximately taking $T_{\rm c}\simeq 0$ in the weak coupling regime, we find that $\Gamma_\alpha(q,i\nu_n\to \omega+i\delta)$ ($\alpha=\parallel,~\perp$) has the form 
\begin{equation}
\Gamma_\alpha={m^2 |k_0| \over 24\pi}{1 \over R_\alpha+i\gamma_\alpha}.
\label{ap.3}
\end{equation}
Here, the real part $R_\alpha$ in the denominator is given by
\begin{equation}
R_\alpha=
\left[
\omega-\left(
\frac{q^2}{2M}-\mu_{\rm B}
\right)
\right]
+F_\alpha+Q,
\label{eq.3.6}
\end{equation}
where 
\begin{eqnarray}
F_{\parallel}(q,\omega)= \frac{4m \omega^2}{\pi q^2|k_0|}
\sqrt{2m\mu}+ \frac{2m^2\omega^3}{\pi q^3|k_0|} 
\log \left|
\frac{q^2+q\sqrt{2m\mu}-2m\omega}{q^2-q\sqrt{2m\mu}-2m\omega}
\right|,
\label{ap.1}
\end{eqnarray}
\begin{eqnarray}
F_{\perp}(q,\omega)&=& -\frac {2m \omega^2}{\pi q^2|k_0|}
\sqrt{2m\mu}- \frac{4m\omega}{\pi q|k_0|} 
\left[
\frac{m\omega^2}{3q^2}+\frac{q^2}{4m}-2\mu-\omega
\right]
\log \left|
\frac{q^2+q\sqrt{2m\mu}-2m\omega}{q^2-q\sqrt{2m\mu}-2m\omega}
\right|,
\nonumber
\\
\label{ap.2}
\end{eqnarray}
\begin{eqnarray}
Q(q,\omega)= 
{24\pi \over m|k_0|}\times
\left\{
\begin{array}{ll}
\frac{|\zeta|^{3/2}}{2}\left(
\log \left| \frac{q-2\sqrt{2m\mu}+2\sqrt{|\zeta|}}{q-2\sqrt{2m\mu}-2\sqrt{|\zeta|}} \right|
+\log \left| \frac{q+2\sqrt{2m\mu}-2\sqrt{|\zeta|}}{q+2\sqrt{2m\mu}+2\sqrt{|\zeta|}} \right|
\right)& ~~~~~\zeta<0,\\
\zeta^{3/2} \left(
\tan^{-1}\frac{2\sqrt{2m\mu}-q}{2\sqrt{\zeta}}+
\tan^{-1}\frac{2\sqrt{2m\mu}+q}{2\sqrt{\zeta}}
\right)& ~~~~~\zeta>0,
\end{array}
\right.
\label{eq.3.8}
\end{eqnarray} 
with $\zeta=q^2/4-2m\mu-m\omega$. The imaginary part $\gamma_\alpha$ in the denominator of Eq. (\ref{ap.3}) is given by
\begin{eqnarray}
\gamma_{\parallel}(q,\omega)
=\frac{2 |\zeta|^{3/2}}{m|k_0|} \eta^3 {\rm sgn}(\omega)\Theta(\zeta),
\label{eq.3.7}
\end{eqnarray}
\begin{eqnarray}
\gamma_\perp(q,\omega)=
\frac{4 |\zeta|^{3/2}}{m|k_0|} \left[3\eta-\eta^3 \right]
{\rm sgn}(\omega)\Theta(\zeta),
\end{eqnarray} 
where $\eta={\rm Min}[1,m|\omega|/(\sqrt{|\zeta|}q)]$. In the present case ($|k_0|/p_{\rm F}\gg 1$), we find from explicit numerical calculations that $F_\alpha$ and $Q$ in Eq. (\ref{eq.3.6}) are negligibly small (although we do not explicitly show the result here), so that we may safely ignore them. The resulting $\Gamma_\alpha$ has the same for as the Bose Green's function with the chemical potential $\mu_{\rm B}$ and with a finite lifetime $\gamma_\alpha^{-1}$, as
\begin{equation}
\Gamma_{\alpha}(q,\omega+i\delta)= 
\frac{24\pi}{\displaystyle m^2 |k_0|}\frac{1}{\displaystyle \omega-\left(\frac{q^2}{2M}-\mu_{\rm B}\right)+ i\gamma_\alpha},
\label{eq.3.9b}
\end{equation}
When the effective range $k_0$ is small, $F_\alpha$ and $Q$ cannot be ignored in Eq. (\ref{eq.3.6}), so that Eq. (\ref{eq.3.9b}) is not obtained.
\par
We note that the molecular excitations at $\omega=q^2/(2M)-\mu_{\rm B}$ in Eq. (\ref{eq.3.9b}) strongly affect single-particle excitations in this regime.  When we consider a Fermi atom with the energy $\omega$ and the momentum ${\bf p}$, such molecular excitations occur when 
\begin{equation}
\omega+\xi_{{\bf p}'}={q^2 \over 2M}-\mu_{\rm B},~~~({\bf p}+{\bf p}'={\bf q})
\label{ap10}
\end{equation}
is satisfied, where $\xi_{{\bf p}'}$ is the energy of another Fermi atom with the momentum ${\bf p}'$ below the Fermi level. In Figs.\ref{fig8}(a) and (b), the region between the two dashed line satisfies this condition. Inside this region, we clearly find the broadening of the spectral peak, which means the short lifetime of the fermionic state due to the coupling with the molecular excitations. In the weak-coupling limit, this region shifts to high energy, so that the ordinary sharp spectral weight is recovered. 
\par
\section{summary}
To summarize, we have discussed strong-coupling effects of a one-component uniform Fermi gas with a tunable $p$-wave pairing interaction. Treating pairing fluctuations in the $p$-wave Cooper channel within the strong-coupling $T$-matrix theory, we have calculated the single-particle density of states (DOS), as well as the spectral weight (SW), in the normal state above $T_{\rm c}$.
\par
Starting from the weak-coupling regime, we showed that the pseudogap gradually develops in DOS and SW, reflecting the enhancement of $p$-wave pairing fluctuations. However, when the pairing interaction becomes strong to some extent, this pseudogap becomes less remarkable to eventually disappear. This result is quite different from the case of the ordinary $s$-wave interaction, where the pseudogap simply becomes large with increasing the interaction strength. We clarified that the difference between the $p$-wave case and $s$-wave case originates from the momentum dependence of the former interaction, having the form $-U{\bf p}\cdot{\bf p}'$. That is, while the coupling constant $U$ become large in the strong-coupling regime, the momentum $p\sim\sqrt{2m\mu}$ ($\equiv{\tilde p}_{\rm F}$) where the pseudogap appears becomes small, due to the decrease of the Fermi chemical potential $\mu$ by the strong-coupling effect. As a result of the competition between the increase of $U$ and the decrease of ${\tilde p}_{\rm F}$, the non-monotonic interaction dependence of the pseudogap phenomenon occurs. 
\par
We determined the pseudogap temperature $T^*$ as the temperature at which the dip structure disappears in DOS. Using this, we identified the pseudogap region in the phase diagram with respect to the temperature and the interaction strength.  
\par
We also showed that molecular excitations still affect single-particle excitations in the weak-coupling regime, when the effective range $k_0$ is much larger than the Fermi momentum $p_{\rm F}$. Since this condition is satisfied in $^{40}$K\cite{Ticknor}, it would be interesting to observe this many-body effect by the photoemission-type experiment developed by JILA group\cite{Gaebler,Stewart}. 
\par
 Since the $p$-wave Fermi superfluid is expected to have richer physics than the conventional $s$-wave state, the realization of this unconventional pairing state would contribute to further development of cold atom physics. Since the pseudogap phenomenon is deeply related to the superfluid phase transition, our results would be useful for research toward the realization of this exciting pairing state in cold Fermi gases.
\par

\acknowledgments
We would like to thank S. Tsuchiya, S. Watabe, and T. Kashimura for useful discussions. This work was supported by Grantin Aid from JSPS, and GCOE Program hHigh-Level Global Cooperation for Leading-Edge Platform on Access Spaces (C12). Y. O. was supported by Grant-in-Aid for Scientific research from MEXT in Japan (22540412, 23104723, 23500056).



\begin{thebibliography}{99}
\bibitem{Regal} C. A. Regal, C. Ticknor, J. L. Bohn, and D. S. Jin, Phys. Rev. Lett. {\bf 90}, 053201 (2003).
\bibitem{Ticknor} C. Ticknor, C. A. Regal, D. S. Jin, and J. L. Bohn, Phys. Rev. A {\bf 69}, 042712 (2004).
\bibitem{Zhang} J. Zhang, E. G. M. van Kempen, T. Bourdel, L. Khaykovich, J. Cubizolles, F. Chevy, M. Teichmann, L. Tarruell, S. J. J. M. F. Kokkelmans, and C. Salomon, Phys. Rev. A {\bf 70}, 030702(R) (2004).
\bibitem{Schunck} C. H. Schunck, M. W. Zwierlein, C. A. Stan, S. M. F. Raupach, W. Ketterle, A. Simoni, E. Tiesinga, C. J. Williams, and P. S. Julienne, Phys. Rev. A {\bf 71}, 045601 (2005).
\bibitem{Ohashi} Y. Ohashi, Phys. Rev. Lett. {\bf 94}, 050403 (2005).
\bibitem{Ho} T. Ho and R. Diener, Phys. Rev. Lett. {\bf 94}, 090402 (2005).
\bibitem{Gurarie}V. Gurarie, L. Radzihovsky, and A. V. Andreev, Phys. Rev. Lett. {\bf 94}, 230403 (2005).
\bibitem{Gurarie2}V. Gurarie, L. Radzihovsky,  Ann. Phys. {\bf 322}, 2 (2007).
\bibitem{Levinsen}J. Levinsen, N. R. Cooper, and V. Gurarie, Phys. Rev. Lett. {\bf 99}, 210402 (2007).
\bibitem{Botelho}S. S. Botelho and C. A. R. S\'a deMelo, J. Low Temp. Phys. {\bf 140}, 409 (2005).
\bibitem{Iskin}M. Iskin and C. A. R. S\'a de Melo, Phys. Rev B {\bf 72}, 224513 (2005).
\bibitem{Iskin2}M. Iskin and C. A. R. S\'a de Melo, Phys. Rev. Lett. {\bf 96}, 040402 (2006).
\bibitem{Iskin3}M. Iskin and C. J.Williams, Phys. Rev. A {\bf 77}, 041607(R) (2008).
\bibitem{Grosfeld}E. Grosfeld, N. R. Cooper, A. Stern, and R. Ilan Phys. Rev. B {\bf 76}, 104516 (2007).
\bibitem{Mizushima}T. Mizushima, M. Ichioka, and K. Machida, Phys. Rev. Lett. {\bf 101}, 150409 (2008).
\bibitem{Mizushima2}T. Mizushima and K. Machida, Phys. Rev. A {\bf 81}, 053605 (2010).
\bibitem{Mizushima3}T. Mizushima and K. Machida Phys. Rev. A {\bf 82}, 023624 (2010). 
\bibitem{Han}Y.-J. Han, Y.-H. Chan, W. Yi, A. J. Daley, S. Diehl, P. Zoller, and L.-M. Duan Phys. Rev. Lett. {\bf 103}, 070404 (2009).
\bibitem{Cheng}C.-H. Cheng and S.-K. Yip, Phys. Rev. Lett. {\bf 95}, 070404 (2005).
\bibitem{Cheng2}C.-H. Cheng and S.-K. Yip, Phys. Rev. B {\bf 73}, 064517 (2006).
\bibitem{Maier}R. A. W. Maier, C. Marzok, and C. Zimmermann, and Ph. W. Courteille, Phys. Rev. A {\bf 81}, 064701 (2010).
\bibitem{Gunter}K. G\"unter, T. St\"oferle, H. Moritz, M. K\"ohl, and T. Esslinger, Phys. Rev. Lett. {\bf 95}, 230401 (2005).
\bibitem{Regal2} C. A. Regal, C. Ticknor, J. L. Bohn, and D. S. Jin, Nature (London) {\bf 424}, 47 (2003).
\bibitem{Geabler} J. P. Gaebler, J. T. Stewart, J. L. Bohn, and D. S. Jin, Phys. Rev. Lett. {\bf 98}, 200403 (2007).
\bibitem{Inaba}Y. Inada, M. Horikoshi, S. Nakajima, M. Kuwata-Gonokami, M. Ueda, and T. Mukaiyama, Phys. Rev. Lett. {\bf 101}, 100401 (2008).
\bibitem{Fuchs}J. Fuchs  C. Ticknor, P. Dyke, G. Veeravalli, E. Kuhnle, W. Rowlands, P. Hannaford, and C. J. Vale , Phys. Rev. A {\bf 77}, 053616 (2008).
\bibitem{Stewart} J. T. Stewart, C. A. Regal, and D. S. Jin, Nature {\bf 454}, 744 (2008).
\bibitem{Gaebler} J. P. Gaebler, J. T. Stewart, T. E. Drake, D. S. Jin, A. Perali,	P. Pieri, and G. C. Strinati, Nature Physics, \textbf{6}, 569 (2010).
\bibitem{Perali2} A. Perali, F. Palestini, P. Pieri, G. C. Strinati, J. T. Stewart, J. P. Gaebler, T. E. Drake, and D. S. Jin, Phys. Rev. Lett. \textbf{106}, 060402 (2011).
\bibitem{Salomon} S. Nascimbene, N. Navon, S. Pilati, F. Chevy, S. Giorgini, A. Georges, and C. Salomon, Phys. Rev. Lett. {\bf 106}, 215303 (2011).
\bibitem{Levin} Q. Chen, and K. Levin, Phys. Rev. Lett. {\bf 102}, 190402 (2009).
\bibitem{Tsuchiya1} S. Tsuchiya, R. Watanabe, Y. Ohashi, Phys. Rev. A {\bf 80}, 033613 (2009).
\bibitem{Tsuchiya2} S. Tsuchiya, R. Watanabe, Y. Ohashi, Phys. Rev. A {\bf 82}, 033629 (2010).
\bibitem{Tsuchiya3} S. Tsuchiya, R. Watanabe, Y. Ohashi, Phys. Rev. A {\bf 84}, 043647 (2011).
\bibitem{Hui} H. Hu, X.-J. Liu, P. D. Drummond, and H. Dong, Phys. Rev. Lett. \textbf{104}, 240407 (2010).
\bibitem{Bulgac} P. Magierski, G. Wlazlowski, and A. Bulgac, Phys. Rev. Lett. {\bf 107}, 145304 (2011).
\bibitem{Perali}A. Perali, P. Pieri, G. C. Strinati, and C. Castellani, Phys. Rev. B {\bf 66}, 024510 (2002).
\bibitem{note1} These three kinds of molecules can be also classified by using the angular momenta $L_z=\pm 1$, and 0.
\bibitem{note2} As shown in Fig.\ref{fig3}(b), $\mu$ becomes negative when $1/(vp_{\rm F}^3)\gesim 0$. In Fig.\ref{fig5}, $\mu/\varepsilon_{\rm F}=0.00$ when $1/(vp_{\rm F}^3)=0$, and $\mu/\varepsilon_{\rm F}=-0.27$ when $1/(vp_{\rm F}^3)=4$.
\bibitem{Perali3} A. Perali, P/ Pieri, G. C. Strinati, and C. Castellani, Phys. Rev. B {\bf 66}, 024510 (2002).
\bibitem{Vollhardt} See, for example, D. Vollhardt, and P. W\"olfle, {\it The Superfluid Phases of Helium 3} (Taylor and Frances, London, 1990).
\end{thebibliography}
\end{document}